\documentclass[11pt,a4paper]{article}
\usepackage[T1]{fontenc}
\usepackage{mathptmx} 
\usepackage[pdftex]{graphicx} 
\usepackage[english]{babel} 
\usepackage[pdftex,linkcolor=black,pdfborder={0 0 0}]{hyperref} 
\usepackage{calc} 
\usepackage{enumitem} 
\usepackage{setspace}
\usepackage[a4paper, lmargin=0.12\paperwidth, rmargin=0.12\paperwidth, tmargin=0.1111\paperheight, bmargin=0.1111\paperheight]{geometry} 

\usepackage[all]{nowidow} 
\usepackage[protrusion=true,expansion=true]{microtype} 
\usepackage{csquotes}
\usepackage[affil-it]{authblk}
\usepackage{xcolor}

\usepackage[backend=biber,style=chem-acs, autocite=superscript]{biblatex} 
\newcommand{\mat}[1]{$\mathrm{#1}$} 
\newcommand{\cm}{$\mathrm{cm^{-1}}$} 
\newcommand{\cms}{$\mathrm{cm^{-1}}\ $} 

\author[1,2,*]{Nicola Melchioni}
\author[1,2]{Giacomo Trupiano}
\author[2,3]{Giorgio Tofani}
\author[2]{Riccardo Bertini}
\author[3]{Andrea Mezzetta}
\author[1]{Federica Bianco}
\author[3]{Lorenzo Guazzelli}
\author[1]{Fabio Beltram}
\author[3]{Christian Silvio Pomelli}
\author[2]{Stefano Roddaro}
\author[1,2]{Alessandro Tredicucci}
\author[1,4,$\dagger$]{Federico Paolucci}

\affil[1]{NEST Laboratory, Istituto Nanoscienze CNR and Scuola Normale Superiore, Piazza San Silvestro 12, I-56127, Pisa, Italy}
\affil[2]{Dipartimento di Fisica "E. Fermi", Università di Pisa, Largo Bruno Pontecorvo 3, I-56127, Pisa, Italy}
\affil[3]{Dipartimento di Farmacia, Università di Pisa, Via Bonanno 33, 56126, Pisa, Italy}
\affil[4]{INFN Sezione di Pisa, Largo Bruno Pontecorvo 3, I-56127 Pisa, Italy}
\affil[*]{Author to whom the correspondence should be addressed: nicola.melchioni@sns.it}
\affil[$\dagger$]{Electronic mail: federico.paolucci@pi.infn.it}

\title{Optical grade bromide-based thin film electrolytes}
\begin{document}

\maketitle 
\section*{Abstract}
Controlling the charge density in low-dimensional materials with an electrostatic potential is a powerful tool to explore and influence their electronic and optical properties. Conventional solid gates impose strict geometrical constraints to the devices and often absorb electromagnetic radiation in the infrared (IR) region. A powerful alternative is ionic liquid (IL) gating. This technique only needs a metallic electrode in contact with the IL and the highest achievable electric field is limited by the electrochemical interactions of the IL with the environment. Despite the excellent gating properties, a large number of ILs is hardly exploitable for optical experiments in the mid-IR region, because they typically suffer from low optical transparency and degradation in ambient conditions. Here, we report the realization of two electrolytes based on bromide ILs dissolved in polymethyl methacrylate (PMMA). We demonstrate that such electrolytes can induce state-of-the-art charge densities as high as $20\times10^{15}\ \mathrm{cm^{-2}}$. Thanks to the low water absorption of PMMA, they work both in vacuum and in ambient atmosphere after a simple vacuum curing.
Furthermore, our electrolytes can be spin coated into flat thin films with optical transparency in the range from 600 cm$^{-1}$ to 4000 cm$^{-1}$. Thanks to these properties, the electrolytes are excellent candidates to fill the gap as versatile gating layers for electronic and mid-IR optoelectronic devices.

\newpage

The possibility of controlling the charge density in an electron gas with electrostatic potentials (field-effect) unlocked many technologies at the basis of modern life. Such field-effect found new importance in experiments involving low-dimensional systems, such as 2D materials, Van der Waals heterostructures and nanowires. In those materials, a modification of the electronic density can have great impact on charge transport and infrared (IR) optical response \cite{yan_2011, grigorenko_2012}. Moreover, field-effect dramatically influences the band structure of 2D crystals \cite{oostinga_2008, zhang_2009, xianqi_2015, kim_2015, forsythe_2018, peng2019}, since their few-atomic thickness prevents an efficient screening of the electric field created by the gate electrode. 
Conventionally, different solid-gate geometries are exploited, such as highly doped substrates covered with a dielectric (back-gate \cite{lemme_2007}), a dielectric deposited on the crystal hosting an electrode on top (top-gate \cite{zhang_2009}), lateral metallic pads on the same insulating substrate (side-gate \cite{Robinson_2003}) or even Van der Waals heterostructures, where different layers act as gate electrode, dielectric and channel medium \cite{novoselov_2016}. 
When dealing with optical experiments involving 2D materials and bulk semiconductors, such geometries are a limitation. 
Indeed, the high dielectric losses hinder the back-gating efficiency of the substrate, a top-gate would prevent adequate optical transparency, while side-gates do not grant the necessary spatial uniformity in the achievable charge density.

In this framework, ionic liquids (ILs) could be a versatile tool to overcome these restrictions. Indeed, ILs work independently from the surrounding environment and need only to be electrically contacted, thus being employable on fully insulating substrates and several gating geometries \cite{chenguang_2004, rajiv_2007, segawa_2012, Piatti_2021}. In the presence of a voltage between the channel and the gate electrode, the ions inside the IL arrange as charged layers at their surfaces.
The electrostatic potential created by these ionic layers is screened by the attraction of charge carriers (electrons or holes) inside the channel and the gate electrode, thus forming the so-called electric double layers (EDLs) \cite{fujimoto_2013}. These EDLs are parallel plate capacitors with inter-plate distance of a few nm, thus offering an enhanced gating efficiency due to their high capacitance \cite{zhang_2019, Gonnelli_2017,Daghero_2012}. 
The generated electric fields up to 10 GV/m \cite{weintrub_2022} allowed to explore unprecedented regimes, where deep modifications to the electronic properties were induced \cite{ye_2012, wang_2012, Ponomarev_2018, domaretskiy_2022}. Moreover, while the dielectric breakdown restricts the efficiency of solid-state gates \cite{murrell_1993, hattori_2016}, the main limitation of electrolyte gating is the electrochemical interaction of the IL with the surrounding environment \cite{shishun_2017}.  
For example, despite their high ionic transport efficiency, lithium-based electrolytes are expensive, highly reactive with most of other chemical substances\cite{Sheng_2022} and scarcely compatible with polymer matrices. Differently, bromide-based ILs are widely available at accessible costs, polymer compatible, and at the same time thermally and electrochemically stable\cite{ferdeghini_2021, VRANES_2019}. However, most commercially available compounds are unstable under high voltage bias and exhibit poor transparency in the IR region \cite{PIATTI_2022, palumbo_2020, MOUMENE_2014}.\\
\begin{figure*}[!t]
    \centering
    \includegraphics[width=.95\textwidth]{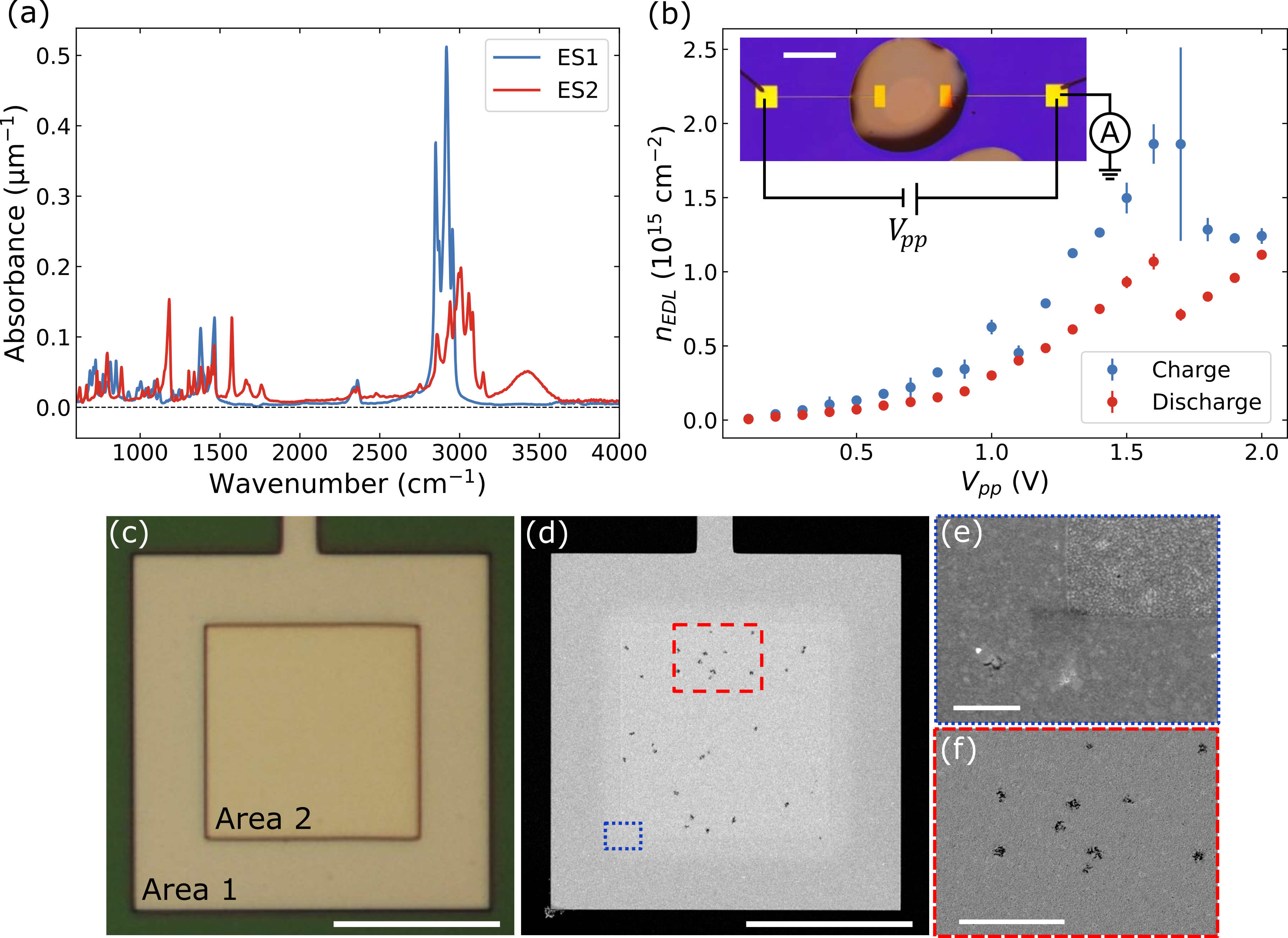}
    \caption{(a) Absorbance spectra of ES1 (blue solid line) and ES2 (red dashed line). (b) Charging ($n_{EDL}^{ch}$, blue dots) and discharging ($n_{EDL}^{dch}$, red dots) carrier density accumulated in the EDL in a drop of ES1 versus $V_{pp}$. Inset: optical image of the IL droplet on the metal pads employed for the transport measurements along with the experimental setup. The scalebar is 500 $\mathrm{\mu m}$. (c) Optical image of a pristine Cr/Au electrode. Area 1 is covered with a PMMA film, Area 2 is exposed to air and will be in direct contact with ES1. The scalebar is 100 \mat{\mu m}. (d) SEM image of the same electrode after the cycles of measurement in ambient conditions. The scalebar is 100 \mat{\mu m}. (e) Detail of a region between Area 1 and Area 2 (blue dotted square in (d)). The scalebar is 1 \mat{\mu m}. (f) Detail of Area 2 (red dashed square in (d)). The scalebar is 5 \mat{\mu m}. }
    \label{fig1}
\end{figure*}
In this Letter, we present the development of two new IR-transparent polymer electrolytes based on bromide ionic liquids (br-ILs) dissolved in polymethyl methacrylate (PMMA). To this scope, we study their optical and charge transport properties when they are spin coated onto a substrate to form a thin film. 
In particular, we show that our electrolytes have large windows with $<0.007 \mathrm{\mu m^{-1}}$ absorption in the range from 600 \cms to 4000 \cm in air and are able to induced charge densities up to $\sim 20 \times 10^{15}\ \mathrm{cm^{-2}}$ in vacuum. Since their fabrication does not impose any geometrical restriction, our electrolytes can be employed as versatile transparent gates for low-dimensional materials in optical and optoelectronic experiments in the IR region. In addition, thanks to the flat surface, such electrolytic films might be an easily-accessible key element for a wide variety of experiments, such as light-matter interaction studies on mixed 2D-bulk semiconductor structures or double-gated systems at extreme charge concentrations.\\
We first focus on the IR optical response and transport properties of two pure bromide-based ionic liquids (ES1 and ES2). ES1 is commercial tetraoctylphosphonium bromide [\mat{P(C8)_4Br}] of purity $>95\%$, while ES2 is 3,3'-(hexane-1,5-diyl)bis(1-methyl-1H-imidazolium)bromide [\mat{C_6(MIM)_2/2Br\ 3C6\ Br}]. 
ES2 was synthesized using the standard ionic liquid synthesis procedure \cite{chiappe_2016} adapted for dicationic ionic liquids \cite{ferdeghini_2021}. A solution (2.1 equiv : 5 mL) of 1-methylimidazole ($98\%$) in acetonitrile ($\geq 99\%$) was added dropwise under magnetic stirring to a solution (2 g, 1 equiv) of 1,6-dibromohexane ($98\%$) in 10 mL of acetonitrile. The solution was stirred for 2 min and then the mixture was heated up and stirred at \mat{80\ ^{\circ}C} for 48 h. A white solid precipitation was observed after half an hour. The solvent was removed under reduced pressure and the obtained solid was washed 3 times with 10 mL of diethyl ether ($99\%+$) and dried \textit{in vacuo} at 65 °C.\\
To determine the IR optical properties of ES1 and ES2, two solid samples of the ILs were measured in a FTIR spectroscope with the Attenuated Total Reflection (ATR) method \cite{Mendoza-Galvan_21}. The measured absorbance spectrum of ES1 (blue solid line in Fig. \ref{fig1}a) shows an absorbance generally lower than 0.1 $\mathrm{\mu m^{-1}}$ in the range from 600 \cms to 2500 \cm, except for one sharp feature of $\sim 0.12\ \mathrm{\mu m^{-1}}$ at 1490 \cm. Higher absorbance peaks close to $0.5\ \mathrm{\mu m^{-1}}$ are visible below 3000 \cm. These peaks are attributed to the stretching vibrations of the \mat{C_{sp3}-H_{x=1,2,3}} groups present in tetraoctylphosphonium bromide \cite{silverstein}. Conversely, the absorbance spectrum of ES2 (red dashed line in Fig. \ref{fig1}a) shows more structured peaks up to $0.16\  \mathrm{\mu m^{-1}}$ in the region from 600 \cms to 2500 \cms and peaks around 3000 \cms lower with respect to ES1. These differences stem from the different cationic moiety of ES2, thus from the vibration modes of the imidazolium ring and from the stretching vabriation of the shorter aliphatic chain and of the \mat{C_{sp2}-H}, respectively\cite{yamada_2017}. In both spectra, the peaks around 2350 \cms are due to the absorption of \mat{CO_2} in the atmosphere and are not related to the ILs.
\\
We now focus on the ionic transport properties of ES1. To this scope, we fabricated metallic electrodes on top of boron-doped \mat{Si} substrates covered with 300 nm of thermally grown \mat{SiO_2} by electron beam lithography (EBL) and thermal evaporation of 7 nm of chromium and 50 nm of gold (see the inset of Fig. \ref{fig1}b). 
The transport properties of pure ES1 were investigated by Double Step Chronocoulometry\cite{Inzelt2010} (DSC, see Fig. S2 in SI) in ambient conditions.
Thus, the accumulated charge in the EDL ($Q_{EDL}$) were determined both during its charging ($ch$) and discharging ($dch$). Then, the values of $Q_{EDL}$ were divided by the electrode area and by the electronic charge to extract the charge carrier density accumulated in (removed from) the EDL, i.e. $n_{EDL}^{ch}$ ($n_{EDL}^{dch}$). 
Figure \ref{fig1}b reports $n_{EDL}^{ch}$ and $n_{EDL}^{dch}$ as a function of the applied potential ($V_{pp}$) for ES1. The accumulated charge density grows with $V_{pp}$ until reaching a value of $\sim1.8\times 10^{15}$ \mat{cm^{-2}} for $V_{pp} = 1.6$ V. For larger $V_{pp}$, the accumulated charge decreases. Such behaviour is the result of the electrochemical activity of the system. Indeed, the upper limit of the electrochemical window \cite{Hayyan_2013} of ES1 is $\sim 1.25$ V (see Fig S3 in SI). When the applied voltage exceeds such value, reduction/oxidation reactions start to take place at the interface between the liquid and the electrodes \cite{shishun_2017, Girault2004}, effectively reducing the charge accumulated in the EDL. Moreover, water from the atmosphere can dissolve in ES1. When applying a voltage bias, the absorbed water reacts with the bromide in the IL forming an acid environment \cite{shkrob_2013}. Consequently, part of the \mat{Br^-} anions involved in these reactions do not participate to the charging of the EDL.
The electrochemical activity of the liquid generates a difference between the charging and the discharging of the EDL (red dots in Fig. \ref{fig1}b). In particular, $Q_{EDL}^{ch} \sim 1.7 Q_{EDL}^{dch}$ for all values of $V_{pp}<1.6$ V (see Fig. S4 in SI). This asymmetry can be partially linked to the mobility of the ions in the IL, since the ionic mobility depends on the driving field and it changes when a voltage is applied with respect to the zero-bias condition \cite{clark_2019}. However, the largest difference is observed for $V_{pp} = 1.7$ V, just above the electrochemical window limit, and the difference lowers as the voltage is increased. This behaviour suggests that the change in the ionic mobility, which should be increased by the higher applied voltage, has in fact minor effects with respect to the reactions.\\
To better understand the interaction between the liquid and the electrodes, the experiment was repeated on a device with electrodes partially covered by a  PMMA with opening fabricated by EBL (see Fig. \ref{fig1}c). After the transport measurements, ES1 and the PMMA mask were removed with acetone and the electrodes observed under a scanning electron microscope (SEM). The contrast between the two regions shows that the morphology of the metallic film that was in contact with ES1 is dramatically different from the one of the masked metal (see Fig. \ref{fig1}d), thus confirming that the ionic liquid chemically interacted with the electrodes. In addition, a 10-${\mu m}$-wide transitional region is visible, due to the IL diffusion below the PMMA mask (see Fig. \ref{fig1}e). 
In the region exposed to ES1, the overall roughness is increased and holes are present in the metallic film (see Fig. \ref{fig1}f). Such observation suggests that the electrochemically formed bromide-based acids attack the electrodes when applying a voltage. Indeed, the interaction of bromide with water absorbed from the atmosphere can create hypobromous acid (\mat{HOBr}) and hydrobromic acid (\mat{HBr}) \cite{Sivey_2013}. Both acids attack gold \cite{walker1990crc}, thus degrading the electrodes and reducing the gating effectiveness of the liquid. As a consequence, when voltages are applied to the liquid in ambient atmosphere, electrochemical reactions happen also within the electrochemical window. However, when water is removed from the liquid, the electrochemical window widens and such reactions are avoided (see Fig. S3 in SI).\\
Since PMMA can absorb only a small amount of water ($\sim2\%$ w/w)\cite{Diaye_2012}, we exploit it to embed the bromide-based ILs and prevent the creation of an acid environment destructive for the devices. Furthermore, a spin-coated electrolyte thin film improves the device optical properties thanks to its lower thickness and higher flatness. Therefore, we investigated both the optical and charge transport properties of the ILs when embedded in the PMMA matrix.
To this scope we synthesised two polymer electrolytes (PES1 and PES2) starting from ES1 and ES2. 
PES1 is produced by dissolving 4.30 mg of ES1 in 430 mg of AR 679.04 950K (ethyl lactate solution  of PMMA 950K at $4\%$, AllResist), while PES2 is realized by dissolving 6.10 mg of ES2 in 612 mg of AR 679.04 950K. Both solutions are therefore 25\%  mg IL/mg PMMA.\\
\begin{figure}[!t]
    \centering
    \includegraphics[width=.48\textwidth]{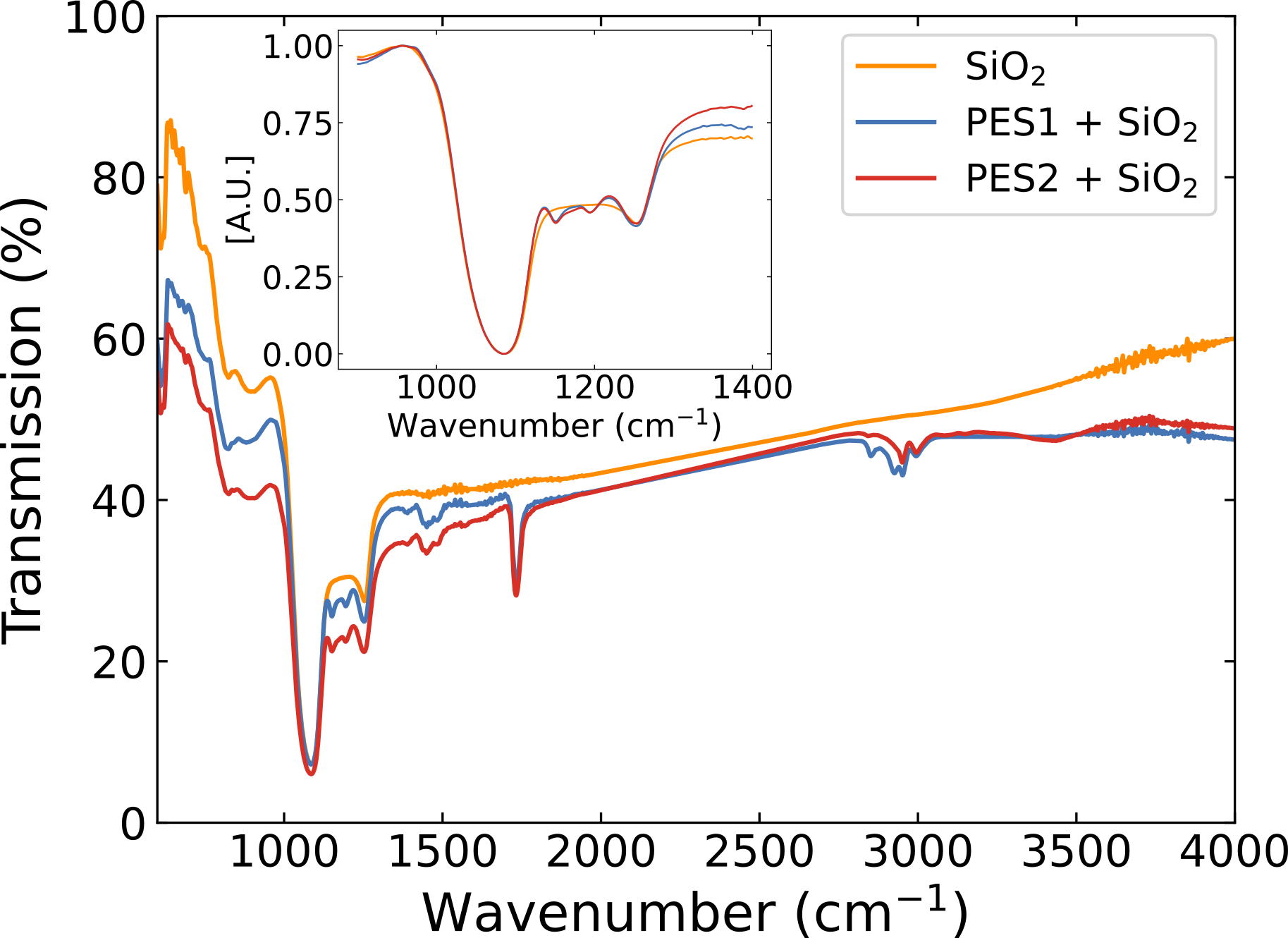}
    \caption{Transmission spectrum of a \mat{Si/SiO_2} chip (orange line) compared to the transmission of the same substrate with PES1 (blue line) and PES2 (red line) on top. Inset: zoom of the signals in the region around the \mat{SiO_2} phonons. The minima were shifted to 0 and the signals normalized on their maxima for a better comparison of the shape.}
    \label{fig2}
\end{figure}
To demonstrate the suitability of PESs in optical experiments, we measured the optical transmission of a structure composed by a PES film spin coated onto \mat{Si/SiO_2} substrates (4000 rpm for 1 minute, soft baked at \mat{100\ ^{\circ}C} for 2 minutes, thickness $\sim\ 290$ nm, see Fig. S1 in SI). 
In general, the PES/\mat{SiO_2/Si} transmission spectra show a reduction of the transmitted power of $\sim 5-10\%$ with respect to the bare substrate (see Fig. \ref{fig2}). This reduction is due to the absorption in the film and to reflection effects caused by the air/PES and PES/\mat{SiO_2} interfaces. The baseline is also influenced by thin film effects \cite{elmenyawy_2013}, visible as a downward bending in the higher region of the spectra. Transmission dips related to PESs are present in the region 1200-1800 \cms. These lines are mainly attributed to the deformation vibrations of the \mat{(O)CH_3} and ester groups in the PMMA film \cite{jitian_2012} and are very similar for both PES1 and PES2. Instead, the two spectra differ in the region close to 3000 \cm, where the resonances are originated by the strain vibration of \mat{CH_{x=1,2,3}} groups \cite{jeon_2008}. Thus, the reduced transmission in the PESs film can be attributed to the different distribution of the \mat{CH_{x=1,2,3}} groups in the films due to the presence of the ESs. Nevertheless, the general features of the transmission spectrum of the bare substrate are preserved, such as the characteristic \mat{SiO_2} phonons at 1084 \cms and 1256 \cms\cite{lehmann_1983} see inset of Fig. \ref{fig2}).\\
Generally, the charge transport properties of the electrolytes are modified when they are embedded in a polymer matrix \cite{mogurampelly2017structure, ganesan_2019,vargas_2020,Xiao_2020, fong_2021} especially when they are in form of a thin film \cite{zhao_2021}. Indeed, the ionic transport is mediated by large molecules that migrate through the polymeric matrix. For example, amorphous polymers offer larger inter-molecular cavities for ions migration with respect to ordered polymers \cite{RAMESH2011}. The dimension of inter-molecular cavities is also influenced by ion size \cite{shen_2020} and by the employed solvent \cite{sharick_2016}. 
To test the suitability of our polymer electrolytes as gating materials, we performed DSC experiments on both PES1 and PES2. In the form of a thin film, PES1 showed a poor ionic transport, thus we focused on PES2. Figure \ref{fig3}a shows the charge density accumulated in the EDL formed by a PES2 film as a function of the applied voltage $V_{pp}$ both in ambient and vacuum conditions. In ambient conditions, the accumulated charge density reaches up to $2\times10^{15}\ \mathrm{cm^{-2}}$ for positive voltages and $4\times10^{15}\ \mathrm{cm^{-2}}$ for negative voltages (blue dots in Fig. \ref{fig3}a). A nominally identical sample was investigated in vacuum at a pressure of $\sim 10^{-5}$ mbar (red dots in Fig. \ref{fig3}a). The accumulated charge density is much higher in vacuum than in air pressure: $n_{EDL}^{vac}/n_{EDL}^{air} \sim 19.7$ for $V_{pp} = 0.5$ V and $\sim 8.6$ for $V_{pp} = -0.5$ V. The higher accumulated charge density in vacuum is attributed to the increased ion mobility as a consequence of two concurring factors. First, when PSEs are in vacuum, contaminant gaseous molecules such as water are expelled from the material. Second, the reduced pressure on the surface allows for the structure to relax, thus creating wider paths for ions to migrate. The asymmetry measured for positive and negative values of the applied potential in vacuum can be attributed to the difference of cations and anions dimension in the liquid. Similarly to pure ES1 (see Fig. \ref{fig1}b), the induced charge density grows following a monotonic trend until $|V_{pp}|\sim 0.9$ V. After that threshold, the ions in the electrolyte start interacting with the polymer matrix and the electrodes, thus modifying the EDL charging efficiency.\\
\begin{figure}[!t]
    \centering
    \includegraphics[width=.48\textwidth]{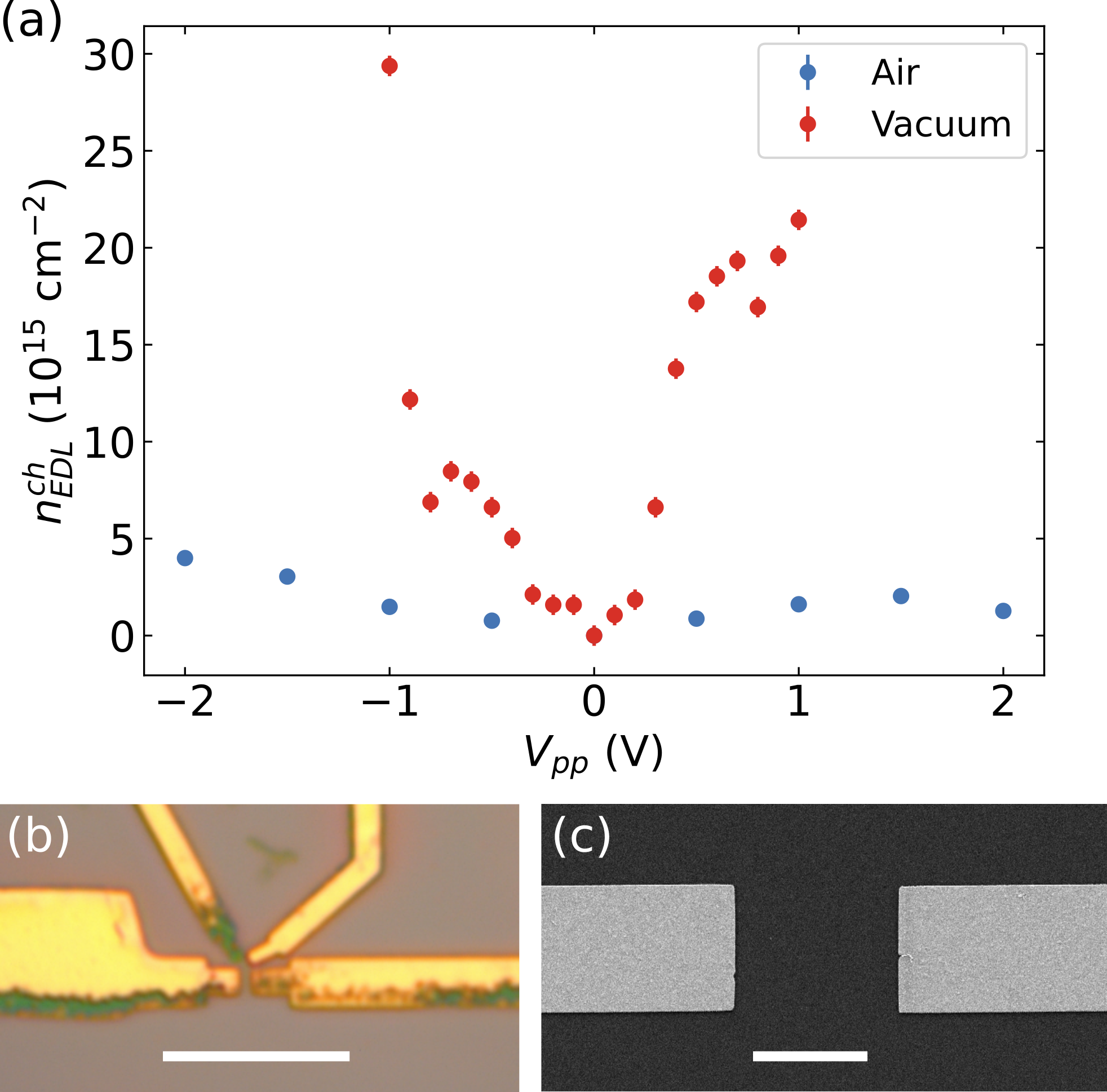}
    \caption{(a) Charge density accumulated in the EDLs in PES2 when in ambient conditions (blue) and vacuum (orange) versus $V_{pp}$. (b) Optical image of the metallic electrodes employed for the measurement in ambient conditions. The scalebar is 5 \mat{\mu m}. (c) SEM image of the metallic electrodes employed for measurement in vacuum. The scalebar is 10 \mat{\mu m}.}
    \label{fig3}
\end{figure}
Despite the IL being embedded in PMMA, an acid environment is created when applying a DC bias to PES2 in ambient conditions, as measured for bare ES1 (see Fig. \ref{fig1}c-f). This acid environment modifies the morphology of the device electrodes, as clearly observable in Fig. \ref{fig3}b. Conversely, when PES2 is measured in vacuum, the electrodes are preserved from electrochemical damaging (Fig. \ref{fig3}c). This indicates the small amount of water trapped by the PMMA film in ambient conditions is sufficient to create HOBr and/or HBr. Instead, in vacuum, the film expels most of the absorbed water, thus preventing the formation of the acid environment.

\begin{figure}[!t]
\centering
    \includegraphics[width=.48\textwidth]{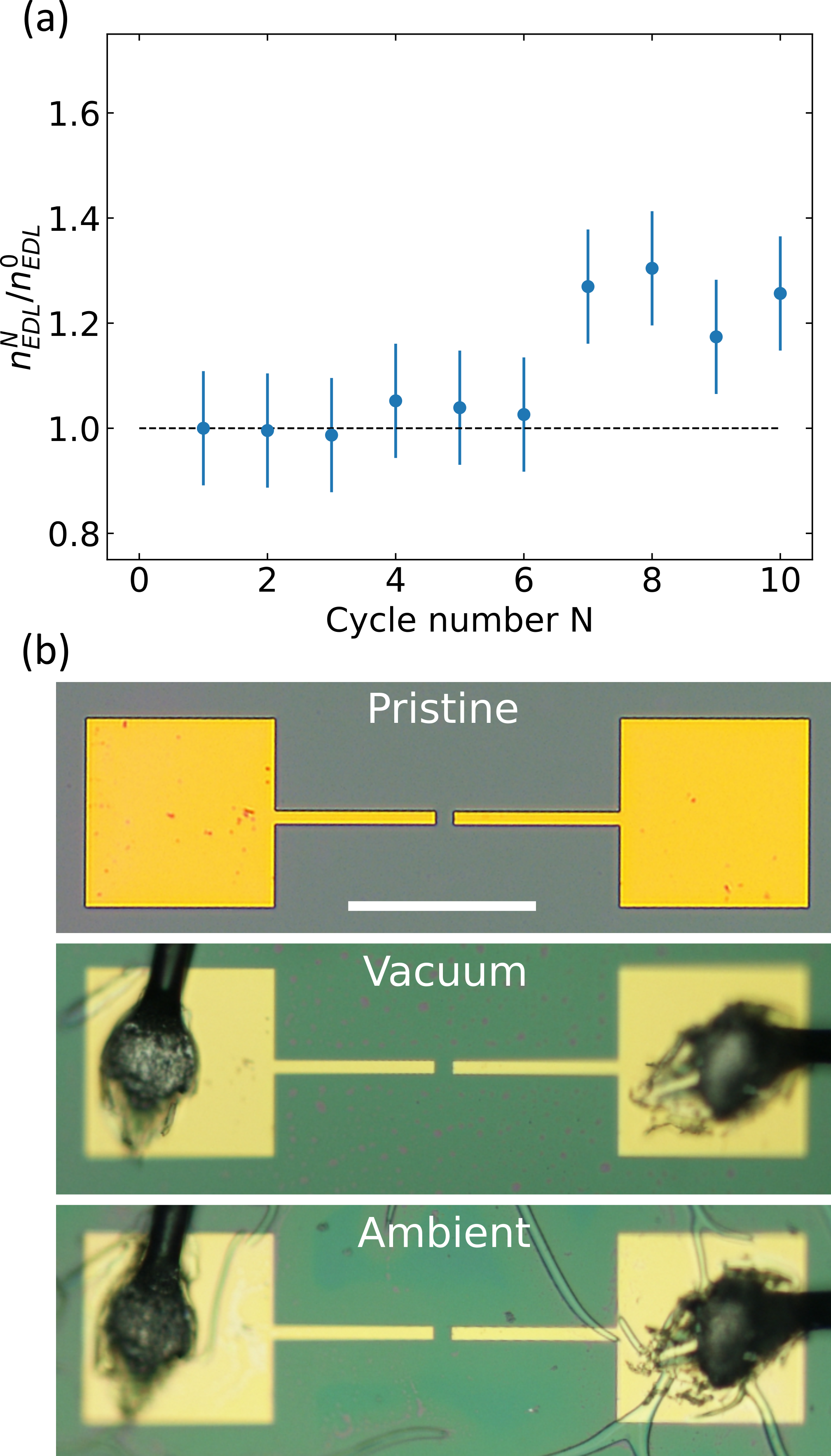}
    \caption{(a) Charge density accumulated in consecutive cycles performed on PES2 in ambient conditions at $V_{pp} = 300$ mV after the measurements in vacuum. (b) Optical images of the metallic pads used in the chronocoulometry measurements before PES2 is spinned (top), after the measurement in vacuum (middle) and after the measurement in air (bottom). The scalebar is 150 \mat{\mu m}. The bonding wires used for electrical measurements are visible on the gold pads in the middle and bottom images.}
    \label{fig4}
\end{figure}

After removing the water by keeping the device in vacuum, we test the stability of the PES structure by performing consecutive DSC cycles with a fixed $V_{pp} = 300$ mV in ambient conditions. As shown in Fig. \ref{fig4}a, the accumulated charge density $n_{EDL}^N$ is highly stable for the first 6 cycles. Here, $n_{EDL}^N$ is the accumulated charge density at cycle $N$ normalized by the carrier concentration accumulated during the first cycle. 
After these cycles, $n_{EDL}^N$ increases of a factor $\sim$ 1.3. This increase can be tentatively explained by considering that, when the polymer is brought back at air conditions, the reduction of polymer film volume caused by the air pressure leads to a decrease of the the paths for the ions migration. Consequently, during the first few cycles, the ions are pushed through the polymer matrix and open back accessible paths for the charge migration. As a result, the charge accumulated in the EDL increases after a few cycles, as similarly occurs for lithium intercalation in bilayer graphene \cite{kuhne_2017, zielinski_2019}. It is worth to notice that, differently from previous experiments, the applied potential does not damage the electrodes, as shown in Fig. \ref{fig4}b. Interestingly, the macroscopic morphology of the PES2 film changes after the application of voltages in ambient conditions after the vacuum treatment. Indeed, bias-induced cracks appear over the surface of the polymer, mainly starting from points where the film was already damaged, such as near the bonding wires or metallic residues, (see Fig. \ref{fig4}b and S5 in SI). Indeed, the water molecules can efficiently penetrate into the PMMA film from these points. Within these cracks, Br- based acids can be electrochemically created and gradually diffuse through the polymer matrix. In full agreement with the DSC measurements, Fig. \ref{fig4}b shows that even when the cracks are formed, the electrodes are not degraded, suggesting that negligible chemical interactions took place. 
Therefore, removing the water content from the polymer matrix before any voltage is applied enhances the robustness of polymer-embedded ionic liquids. Indeed, a simple treatment in vacuum preparation ensures a large stability of the accumulated charge density and prevents the formation of an acid environment in the proximity of the metallic electrodes even when employed for several cycles in air.
In summary, we proposed and demonstrated a new class of polymer-embedded bromide-based ionic liquids. 
The EDLs that form when the films are polarized can accumulate state-of-the-art charge densities up to $\sim 20 \times10^{15}\ \mathrm{cm^{-2}}$ \cite{Piatti_2021}.
Furthermore, these electrolytes are transparent in the mid-infrared region of the spectrum and can be easily spin coated in the form of a thin film, thus providing remarkable advantages in optoelectronic devices. 
Although bromide based ILs react with water to form acids chemically attacking the electrodes, we demonstrated that it is possible to remove the absorbed water by a vacuum treatment before any voltage is applied, thus preventing the degradation of the metals. Furthermore, the polymer matrix of PMMA embedded ILs protects the electrodes for several cycles in ambient conditions after the vacuum curing. Therefore, devices exploiting our electrolytes can be fabricated with standard techniques in ambient conditions without the need of the inert atmosphere of a glove box, as usually necessary for lithium based ILs \cite{kuhne_2017,Daghero_2012,zielinski_2019,Gonnelli_2017}. \\

\section*{Acknowledgments}
Fe.B. acknowledges the project q-LIMA in the framework of the PRIN2020 initiative of the Italian Ministry of University and Research for partial financial support. Gio.T., A.M., L.G., C.S.P. and S.R. acknowledge the project Quantum2D in the framework of the PRIN2017 initiative of the Italian Ministry of University and Research for partial financial support.

\section*{Data Availability}
All the data are available upon request to the corresponding author.

\section*{Conflict of Interests}
The authors have no conflicts to disclose.

\printbibliography

@article{elmenyawy_2013,
title = {Infrared spectra, optical constants and semiconductor behavior of 5-(2-phenylhydrazono)-3,3-dimethylcyclohexanone thin films},
journal = {Journal of Molecular Structure},
volume = {1036},
pages = {144-150},
year = {2013},
issn = {0022-2860},
doi = {10.1016/j.molstruc.2012.09.067},
author = {E.M. El-Menyawy and N.A. El-Ghamaz and H.H. Nawar},
}

@article{lehmann_1983,
author = {Lehmann, A. and Schumann, L. and H{ü}bner, K.},
title = {Optical Phonons in Amorphous Silicon Oxides. I. Calculation of the Density of States and Interpretation of LO-TO Splittings of Amorphous Sio2},
journal = {Physica Status Solidi B},
volume = {117},
number = {2},
pages = {689-698},
doi = {10.1002/pssb.2221170231},
year = {1983}
}

@article{jitian_2012,
author = {Jitian,Simion  and Bratu,Ioan },
title = {Determination of optical constants of polymethyl methacrylate films from IR reflection-absorption spectra},
journal = {AIP Conference Proceedings},
volume = {1425},
number = {1},
pages = {26-29},
year = {2012},
doi = {10.1063/1.3681958},
}

@article{jeon_2008,
author = {Jeon, Yoonnam and Sung, Jaeho and Seo, Choongwon and Lim, Hyunjin and Cheong, Hyeonsik and Kang, Minhyuck and Moon, Bongjin and Ouchi, Yukio and Kim, Doseok},
title = {Structures of Ionic Liquids with Different Anions Studied by Infrared Vibration Spectroscopy},
journal = {The Journal of Physical Chemistry B},
volume = {112},
number = {15},
pages = {4735-4740},
year = {2008},
doi = {10.1021/jp7120752},
URL = { 
        https://doi.org/10.1021/jp7120752
    
},

}

@article{chiappe_2016,
author = {Chiappe, Cinzia and Mezzetta, Andrea and Pomelli, Christian Silvio and Puccini, Monica and Seggiani, Maurizia},
title = {Product as Reaction Solvent: An Unconventional Approach for Ionic Liquid Synthesis},
journal = {Organic Process Research \& Development},
volume = {20},
number = {12},
pages = {2080-2084},
year = {2016},
doi = {10.1021/acs.oprd.6b00302},

}

@article{ferdeghini_2021,
title = {Synthesis, thermal behavior and kinetic study of N-morpholinium dicationic ionic liquids by thermogravimetry},
journal = {Journal of Molecular Liquids},
volume = {332},
pages = {115662},
year = {2021},
issn = {0167-7322},
doi = {https://doi.org/10.1016/j.molliq.2021.115662},
author = {Claudio Ferdeghini and Lorenzo Guazzelli and Christian S. Pomelli and Andrea Ciccioli and Bruno Brunetti and Andrea Mezzetta and Stefano {Vecchio Ciprioti}},
}

@Inbook{Inzelt2010,
author="Inzelt, Gy{\"o}rgy",
editor="Scholz, Fritz
and Bond, A.M.
and Compton, R.G.
and Fiedler, D.A.
and Inzelt, G.
and Kahlert, H.
and Komorsky-Lovri{\'{c}}, {\v{S}}.
and Lohse, H.
and Lovri{\'{c}}, M.
and Marken, F.
and Neudeck, A.
and Retter, U.
and Scholz, F.
and Stojek, Z.",
title="Chronocoulometry",
bookTitle="Electroanalytical Methods: Guide to Experiments and Applications",
year="2010",
publisher="Springer Berlin Heidelberg",
address="Berlin, Heidelberg",
pages="147--158",
isbn="978-3-642-02915-8",
doi="10.1007/978-3-642-02915-8_7",
url="https://doi.org/10.1007/978-3-642-02915-8_7"
}

@article{clark_2019,
author = {Clark,Ryan  and von Domaros,Michael  and McIntosh,Alastair J. S.  and Luzar,Alenka  and Kirchner,Barbara  and Welton,Tom },
title = {Effect of an external electric field on the dynamics and intramolecular structures of ions in an ionic liquid},
journal = {The Journal of Chemical Physics},
volume = {151},
number = {16},
pages = {164503},
year = {2019},
doi = {10.1063/1.5129367},

}

@article{Xiao_2020,
author={Xiao, Wangchuan and Yang, Quan and Zhu, Shenlin},
title={Comparing ion transport in ionic liquids and polymerized ionic liquids},
journal = {Sci. Rep.},
volume = {10},
number = {7825},
year = {2020},
doi = {10.1038/s41598-020-64689-8},
}

@article{zhao_2021,
author = {Zhao, Qiujie and Bennington, Peter and Nealey, Paul F. and Patel, Shrayesh N. and Evans, Christopher M.},
title = {Ion Specific, Thin Film Confinement Effects on Conductivity in Polymerized Ionic Liquids},
journal = {Macromolecules},
volume = {54},
number = {22},
pages = {10520-10528},
year = {2021},
doi = {10.1021/acs.macromol.1c01820},
}

@article{sharick_2016,
author = {Sharick, Sharon and Koski, Jason and Riggleman, Robert A. and Winey, Karen I.},
title = {Isolating the Effect of Molecular Weight on Ion Transport of Non-Ionic Diblock Copolymer/Ionic Liquid Mixtures},
journal = {Macromolecules},
volume = {49},
number = {6},
pages = {2245-2256},
year = {2016},
doi = {10.1021/acs.macromol.5b02445},
}

@article{RAMESH2011,
title = {Evaluation and investigation on the effect of ionic liquid onto PMMA-PVC gel polymer blend electrolytes},
journal = {Journal of Non-Crystalline Solids},
volume = {357},
number = {10},
pages = {2132-2138},
year = {2011},
issn = {0022-3093},
doi = {10.1016/j.jnoncrysol.2011.03.004},
author = {S. Ramesh and Chiam-Wen Liew and K. Ramesh},
}

@article{shen_2020,
author = {Shen, Kuan-Hsuan and Hall, Lisa M.},
title = {Effects of Ion Size and Dielectric Constant on Ion Transport and Transference Number in Polymer Electrolytes},
journal = {Macromolecules},
volume = {53},
number = {22},
pages = {10086-10096},
year = {2020},
doi = {10.1021/acs.macromol.0c02161},
}

@article{fong_2021,
author = {Fong, Kara D. and Self, Julian and McCloskey, Bryan D. and Persson, Kristin A.},
title = {Ion Correlations and Their Impact on Transport in Polymer-Based Electrolytes},
journal = {Macromolecules},
volume = {54},
number = {6},
pages = {2575-2591},
year = {2021},
doi = {10.1021/acs.macromol.0c02545},
}

@article{mogurampelly2017structure,
  title={Structure and mechanisms underlying ion transport in ternary polymer electrolytes containing ionic liquids},
  author={Mogurampelly, Santosh and Ganesan, Venkat},
  journal={The Journal of Chemical Physics},
  volume={146},
  number={7},
  pages={074902},
  year={2017},
  publisher={AIP Publishing LLC}
}

@article{vargas_2020,
author = {Vargas-Barbosa, N. M. and Roling, B.},
title = {Dynamic Ion Correlations in Solid and Liquid Electrolytes: How Do They Affect Charge and Mass Transport?},
journal = {ChemElectroChem},
volume = {7},
number = {2},
pages = {367-385},
doi = {10.1002/celc.201901627},
year = {2020}
}

@Article{ganesan_2019,
author ="Ganesan, Venkat",
title  ="Ion transport in polymeric ionic liquids: recent developments and open questions",
journal  ="Mol. Syst. Des. Eng.",
year  ="2019",
volume  ="4",
issue  ="2",
pages  ="280-293",
publisher  ="The Royal Society of Chemistry",
doi  ="10.1039/C8ME00114F",
}

@book{walker1990crc,
  title={CRC handbook of metal etchants},
  author={Walker, Perrin and Tarn, William H},
  year={1990},
  publisher={CRC press}
}

@article{Hayyan_2013,
title = {Investigating the electrochemical windows of ionic liquids},
journal = {Journal of Industrial and Engineering Chemistry},
volume = {19},
number = {1},
pages = {106-112},
year = {2013},
issn = {1226-086X},
doi = {10.1016/j.jiec.2012.07.011},
author = {Maan Hayyan and Farouq S. Mjalli and Mohd Ali Hashim and Inas M. AlNashef and Tan Xue Mei},
}

@article{shishun_2017,
author = {Zhao, Shishun and Zhou, Ziyao and Peng, Bin and Zhu, Mingmin and Feng, Mengmeng and Yang, Qu and Yan, Yuan and Ren, Wei and Ye, Zuo-Guang and Liu, Yaohua and Liu, Ming},
title = {Quantitative Determination on Ionic-Liquid-Gating Control of Interfacial Magnetism},
journal = {Advanced Materials},
volume = {29},
number = {17},
pages = {1606478},
doi = {10.1002/adma.201606478},
year = {2017}
}

@book{Girault2004,
author={Hubert H. Girault},
publisher={EPFL Press},
title = {Analytical and Physical Electrochemistry},
edition ={1st Edition},
doi = {10.1201/9781439807842},
year ={2004},
}

@article{kuhne_2017,
author = {K{ü}hne, M. and Paolucci, F. and Popovic, J. and Ostrovsky, P.M. and Maier, J and Smet, J.H.},
title = {Ultrafast lithium diffusion in bilayer graphene},
journal = {Nature Nanotechnology},
volume = {12},
pages = {895-900},
year = {2017},
doi = {10.1038/nnano.2017.108},
}

@article{zielinski_2019,
author = {Zielinski, Patrik and K{ü}hne, Matthias and K{ä}rcher, Daniel and Paolucci, Federico and Wochner, Peter and Fecher, Sven and Drnec, Jakub and Felici, Roberto and Smet, Jurgen H.},
title = {Probing Exfoliated Graphene Layers and Their Lithiation with Microfocused X-rays},
journal = {Nano Letters},
volume = {19},
number = {6},
pages = {3634-3640},
year = {2019},
doi = {10.1021/acs.nanolett.9b00654},
}

@article{oostinga_2008,
author = {Oostinga, Jeroen B. and Heersche, Hubert B. and Liu, Xinglan and Morpurgo, Alberto F. and Vandersypen, Lieven M. K.},
title = {Gate-induced insulating state in bilayer graphene devices},
journal = {Nature Materials},
volume = {7},
pages = {151-157},
year = {2008},
doi = {10.1038/nmat2082},

}

@article{zhang_2009,
author = {Zhang, Yuanbo and Tang, Tsung-Ta and Girit, Caglar and Hao, Zhao and Martin, Michael C. and Zettl, Alex and Crommie, Michael F. and Shen, Ron and Wang, Feng},
title = {Direct observation of a widely tunable bandgap in bilayer graphene},
journal = {Nature},
volume = {459},
pages = {820-823},
year = {2009},
doi = {10.1038/nature08105},
}

@article{peng2019,
author = {Chen,Peng  and Cheng,Cai  and Shen,Cheng  and Zhang,Jing  and Wu,Shuang  and Lu,Xiaobo  and Wang,Shuopei  and Du,Luojun  and Watanabe,Kenji  and Taniguchi,Takashi  and Sun,Jiatao  and Yang,Rong  and Shi,Dongxia  and Liu,Kaihui  and Meng,Sheng  and Zhang,Guangyu },
title = {Band evolution of two-dimensional transition metal dichalcogenides under electric fields},
journal = {Applied Physics Letters},
volume = {115},
number = {8},
pages = {083104},
year = {2019},
doi = {10.1063/1.5093055},
}

@article{xianqi_2015,
author = {Dai,Xianqi  and Li,Wei  and Wang,Tianxing  and Wang,Xiaolong  and Zhai,Caiyun },
title = {Bandstructure modulation of two-dimensional WSe2 by electric field},
journal = {Journal of Applied Physics},
volume = {117},
number = {8},
pages = {084310},
year = {2015},
doi = {10.1063/1.4907315},

}

@article{kim_2015,
author = {Jimin Kim  and Seung Su Baik  and Sae Hee Ryu  and Yeongsup Sohn  and Soohyung Park  and Byeong-Gyu Park  and Jonathan Denlinger  and Yeonjin Yi  and Hyoung Joon Choi  and Keun Su Kim },
title = {Observation of tunable band gap and anisotropic Dirac semimetal state in black phosphorus},
journal = {Science},
volume = {349},
number = {6249},
pages = {723-726},
year = {2015},
doi = {10.1126/science.aaa6486},
}

@article{forsythe_2018,
author = {Forsythe, Carlos and Zhou, Xiaodong and Watanabe, Kenji and Taniguchi, Takashi and Pasupathy, Abhay and Moon, Pilkyung and Koshin, Mikito and Kim, Philip and Dean, Cory R.},
title = {Band structure engineering of 2D materials using patterned dielectric superlattices},
journal = {Nature Nanotechnology},
volume = {13},
pages = {566-571},
year = {2018},
doi = {10.1038/s41565-018-0138-7},
}

@article{Robinson_2003,
doi = {10.1088/0957-4484/14/2/336},
url = {https://dx.doi.org/10.1088/0957-4484/14/2/336},
year = {2003},
month = {jan},
publisher = {},
volume = {14},
number = {2},
pages = {290},
author = {L A W Robinson and  S-B Lee and  K B K Teo and  M Chhowalla and  G A J Amaratunga and  W I Milne and  D A Williams and  D G Hasko and  H Ahmed},
title = {Fabrication of self-aligned side gates to carbon nanotubes},
journal = {Nanotechnology},
}

@ARTICLE{lemme_2007,
  author={Lemme, Max C. and Echtermeyer, Tim J. and Baus, Matthias and Kurz, Heinrich},
  journal={IEEE Electron Device Letters}, 
  title={A Graphene Field-Effect Device}, 
  year={2007},
  volume={28},
  number={4},
  pages={282-284},
  doi={10.1109/LED.2007.891668}}

@article{novoselov_2016,
author = {K. S. Novoselov  and A. Mishchenko  and A. Carvalho  and A. H. Castro Neto },
title = {2D materials and van der Waals heterostructures},
journal = {Science},
volume = {353},
number = {6298},
pages = {aac9439},
year = {2016},
doi = {10.1126/science.aac9439},
}

@Article{fujimoto_2013,
author ="Fujimoto, Takuya and Awaga, Kunio",
title  ="Electric-double-layer field-effect transistors with ionic liquids",
journal  ="Phys. Chem. Chem. Phys.",
year  ="2013",
volume  ="15",
issue  ="23",
pages  ="8983-9006",
publisher  ="The Royal Society of Chemistry",
doi  ="10.1039/C3CP50755F",}

@article{chenguang_2004,
author = {Lu, Chenguang and Fu, Qiang and Huang, Shaoming and Liu, Jie},
title = {Polymer Electrolyte-Gated Carbon Nanotube Field-Effect Transistor},
journal = {Nano Letters},
volume = {4},
number = {4},
pages = {623-627},
year = {2004},
doi = {10.1021/nl049937e},

}

@article{Piatti_2021,
doi = {10.1088/2632-959X/ac011d},
year = {2021},
month = {may},
publisher = {IOP Publishing},
volume = {2},
number = {2},
pages = {024003},
author = {Erik Piatti},
title = {Ionic gating in metallic superconductors: A brief review},
journal = {Nano Express},
}

@article{segawa_2012,
  title = {Ambipolar transport in bulk crystals of a topological insulator by gating with ionic liquid},
  author = {Segawa, Kouji and Ren, Zhi and Sasaki, Satoshi and Tsuda, Tetsuya and Kuwabata, Susumu and Ando, Yoichi},
  journal = {Phys. Rev. B},
  volume = {86},
  issue = {7},
  pages = {075306},
  numpages = {7},
  year = {2012},
  month = {Aug},
  publisher = {American Physical Society},
  doi = {10.1103/PhysRevB.86.075306},
}

@article{rajiv_2007,
author = {Misra,Rajiv  and McCarthy,Mitchell  and Hebard,Arthur F. },
title = {Electric field gating with ionic liquids},
journal = {Applied Physics Letters},
volume = {90},
number = {5},
pages = {052905},
year = {2007},
doi = {10.1063/1.2437663},
}

@article{weintrub_2022,
author = {Weintrub, Benjamin I. and Hsieh, Yu-Ling and Kovalchuk, Sviatoslav and Kirchhof, Jan N. and Greben, Kyrylo and Bolotin, Kirill I.},
title = {Generating intense electric fields in 2D materials by dual ionic gating},
journal = {Nature Communications},
volume = {13},
pages = {6601},
year = {2022},
doi = {10.1038/s41467-022-34158-z},
}

@article{zhang_2019,
author = {Zhang, Haijing and Berthod, Christophe and Berger, Helmuth and Giamarchi, Thierry and Morpurgo, Alberto F.},
title = {Band Filling and Cross Quantum Capacitance in Ion-Gated Semiconducting Transition Metal Dichalcogenide Monolayers},
journal = {Nano Letters},
volume = {19},
number = {12},
pages = {8836-8845},
year = {2019},
doi = {10.1021/acs.nanolett.9b03667},

}

@article{murrell_1993,
author = {Murrell,M. P.  and Welland,M. E.  and O’Shea,S. J.  and Wong,T. M. H.  and Barnes,J. R.  and McKinnon,A. W.  and Heyns,M.  and Verhaverbeke,S. },
title = {Spatially resolved electrical measurements of SiO2 gate oxides using atomic force microscopy},
journal = {Applied Physics Letters},
volume = {62},
number = {7},
pages = {786-788},
year = {1993},
doi = {10.1063/1.108579},

}

@article{hattori_2016,
author = {Hattori, Yoshiaki and Taniguchi, Takashi and Watanabe, Kenji and Nagashio, Kosuke},
title = {Anisotropic Dielectric Breakdown Strength of Single Crystal Hexagonal Boron Nitride},
journal = {ACS Applied Materials \& Interfaces},
volume = {8},
number = {41},
pages = {27877-27884},
year = {2016},
doi = {10.1021/acsami.6b06425},

}

@article{ye_2012,
author = {J. T. Ye  and Y. J. Zhang  and R. Akashi  and M. S. Bahramy  and R. Arita  and Y. Iwasa },
title = {Superconducting Dome in a Gate-Tuned Band Insulator},
journal = {Science},
volume = {338},
number = {6111},
pages = {1193-1196},
year = {2012},
doi = {10.1126/science.1228006},}

@article{Ponomarev_2018,
author = {Ponomarev, Evgeniy and Ubrig, Nicolas and Gutiérrez-Lezama, Ignacio and Berger, Helmuth and Morpurgo, Alberto F.},
title = {Semiconducting van der Waals Interfaces as Artificial Semiconductors},
journal = {Nano Letters},
volume = {18},
number = {8},
pages = {5146-5152},
year = {2018},
doi = {10.1021/acs.nanolett.8b02066},}

@article{domaretskiy_2022,
author = {Domaretskiy, Daniil and Philippi, Marc and Gibertini, Marco and Ubrig, Nicolas and Guti{é}rrez-Lezama, Ignacio and Morpurgo, Alberto F.},
title = {Quenching the bandgap of two-dimensional semiconductors with a perpendicular electric field},
journal = {Nature Nanotechnology},
volume = {17},
pages = {1078-1083},
year = {2022},
doi = {10.1038/s41565-022-01183-4},}

@article{wang_2012,
author = {Wang, Shun and Ha, Mingjing and Manno, Michael and Frisbie, C. Daniel and Leighton, C.},
title = {Hopping transport and the Hall effect near the insulator–metal transition in electrochemically gated poly(3-hexylthiophene) transistors},
journal = {Nature Communications},
volume = {3},
number = {1210},
year = {2012},
doi = {10.1038/ncomms2213},}

@article{Gonnelli_2017,
doi = {10.1088/2053-1583/aa5afe},
year = {2017},
month = {jul},
publisher = {IOP Publishing},
volume = {4},
number = {3},
pages = {035006},
author = {R S Gonnelli and E Piatti and A Sola and M Tortello and F Dolcini and S Galasso and J R Nair and C Gerbaldi and E Cappelluti and M Bruna and A C Ferrari},
title = {Weak localization in electric-double-layer gated few-layer graphene},
journal = {2D Materials},
}

@article{yan_2011,
author = {Yan, Hugen and Xia, Fengnian and Zhu, Wenjuan and Freitag, Marcus and Dimitrakopoulos, Christos and Bol, Ageeth A. and Tulevski, George and Avouris, Phaedon},
title = {Infrared Spectroscopy of Wafer-Scale Graphene},
journal = {ACS Nano},
volume = {5},
number = {12},
pages = {9854-9860},
year = {2011},
doi = {10.1021/nn203506n},
}

@article{grigorenko_2012,
author = {Grigorenko, A.N. and Polini, M. and Novoselov, K.},
title = {Graphene Plasmonics},
journal = {Nature Photonics},
volume = {6},
pages = {749-758},
year = {2012},
doi = {10.1038/nphoton.2012.262},
}

@article{Daghero_2012,
author = {Daghero, D. and Paolucci, F. and Sola, A. and Tortello, M. and Ummarino, G. A. and Agosto, M. and Gonnelli, R. S. and Nair, J. R. and Gerbaldi, C.},
title = {Large conductance modulation of gold thin films by huge charge injection via electrochemical gating},
journal = {Physical Review Letters},
volume = {108},
pages = {066807},
year = {2012},
doi = {10.1103/PhysRevLett.108.066807},
}

@article{Diaye_2012,
author = {{N'Diaye}, M. and Pascaretti-Grizon, F. and Massin, P. and Baslé, M. F. and Chappard, D.},
title = {Water Absorption of Poly(methyl methacrylate) Measured by Vertical Interference Microscopy},
journal = {Langmuir},
volume = {28},
pages = {11609–11614},
year = {2012},
doi = {10.1021/la302260a},
}

@article{Sheng_2022,
author = {Sheng, L. and Wang, Q. and Liu, X. and Cui, H. and Wang, X. and Xu, Y. and Li, Z. and Wang, L. and Chen, Z. and Xu, G. and Wang, J. and Tang, Y. and Amine, K. and Xu, H. and He, X.},
title = {Suppressing electrolyte-lithium metal reactivity via $\mathrm{Li^+}$-desolvation in uniform nano-porous separator},
journal = {Nature Communication},
volume = {13},
pages = {172},
year = {2022},
doi = {10.1038/s41467-021-27841-0},
}

@article{VRANES_2019,
title = {Thermophysical and electrochemical properties of 1–alkyl–3–(3–butenyl)imidazolium bromide ionic liquids},
journal = {The Journal of Chemical Thermodynamics},
volume = {139},
pages = {105871},
year = {2019},
doi = {10.1016/j.jct.2019.07.013},
author = {Milan Vraneš and Snežana Papović and Sanja Rackov and Khalaf Alenezi and Slobodan Gadžurić and Aleksandar Tot and Branka Pilić},
}

@article{PIATTI_2022,
title = {Ionic liquids for electrochemical applications: Correlation between molecular structure and electrochemical stability window},
journal = {Journal of Molecular Liquids},
volume = {364},
pages = {120001},
year = {2022},
doi = {10.1016/j.molliq.2022.120001},
author = {Erik Piatti and Luca Guglielmero and Giorgio Tofani and Andrea Mezzetta and Lorenzo Guazzelli and Felicia D'Andrea and Stefano Roddaro and Christian Silvio Pomelli},
}

@Article{palumbo_2020,
author ="Palumbo, O. and Cimini, A. and Trequattrini, F. and Brubach, J.-B. and Roy, P. and Paolone, A.",
title  ="The infrared spectra of protic ionic liquids: performances of different computational models to predict hydrogen bonds and conformer evolution",
journal  ="Phys. Chem. Chem. Phys.",
year  ="2020",
volume  ="22",
issue  ="14",
pages  ="7497-7506",
publisher  ="The Royal Society of Chemistry",
doi  ="10.1039/D0CP00907E",}

@article{MOUMENE_2014,
title = {Vibrational spectroscopic study of ionic liquids: Comparison between monocationic and dicationic imidazolium ionic liquids},
journal = {Journal of Molecular Structure},
volume = {1065-1066},
pages = {86-92},
year = {2014},
doi = {10.1016/j.molstruc.2014.02.034},
author = {Taqiyeddine Moumene and El Habib Belarbi and Boumediene Haddad and Didier Villemin and Ouissam Abbas and Brahim Khelifa and Serge Bresson},
}

@article{yamada_2017,
author = {Yamada, Toshiki and Tominari, Yukihiro and Tanaka, Shukichi and Mizuno, Maya},
title = {Infrared Spectroscopy of Ionic Liquids Consisting of Imidazolium Cations with Different Alkyl Chain Lengths and Various Halogen or Molecular Anions with and without a Small Amount of Water},
journal = {The Journal of Physical Chemistry B},
volume = {121},
number = {14},
pages = {3121-3129},
year = {2017},
doi = {10.1021/acs.jpcb.7b01429},
}

@book{silverstein,
  title={Spectrometric Identification of Organic Compounds, 8th edition},
  author={Silverstein, Robert M. and Webster, Francis X. and Kiemle, David J. and Bryce, David L.},
  year={2014},
  publisher={Wiley}
}

@article{shkrob_2013,
author = {Shkrob, Ilya A. and Marin, Timothy W. and Crowell, R. A. and Wishart, James F.},
title = {Photo- and Radiation-Chemistry of Halide Anions in Ionic Liquids},
journal = {The Journal of Physical Chemistry A},
volume = {117},
number = {28},
pages = {5742-5756},
year = {2013},
doi = {10.1021/jp4042793},
}

@article{Sivey_2013,
author = {Sivey, John D. and Arey, J. Samuel and Tentscher, Peter R. and Roberts, A. Lynn},
title = {Reactivity of BrCl, Br2, BrOCl, Br2O, and HOBr Toward Dimethenamid in Solutions of Bromide + Aqueous Free Chlorine},
journal = {Environmental Science \& Technology},
volume = {47},
number = {3},
pages = {1330-1338},
year = {2013},
doi = {10.1021/es302730h},
}

@article{Mendoza-Galvan_21,
author = {A. Mendoza-Galv\'{a}n and J. G. M\'{e}ndez-Lara and R. A. Mauricio-S\'{a}nchez and K. J\"{a}rrendahl and H. Arwin},
journal = {Opt. Lett.},
number = {4},
pages = {872--875},
publisher = {Optica Publishing Group},
title = {Effective absorption coefficient and effective thickness in attenuated total reflection spectroscopy},
volume = {46},
month = {Feb},
year = {2021},
doi = {10.1364/OL.418277},
}

\end{document}


\maketitle
\doublespacing

\begin{figure}
    \centering
    \includegraphics[width = .9\textwidth]{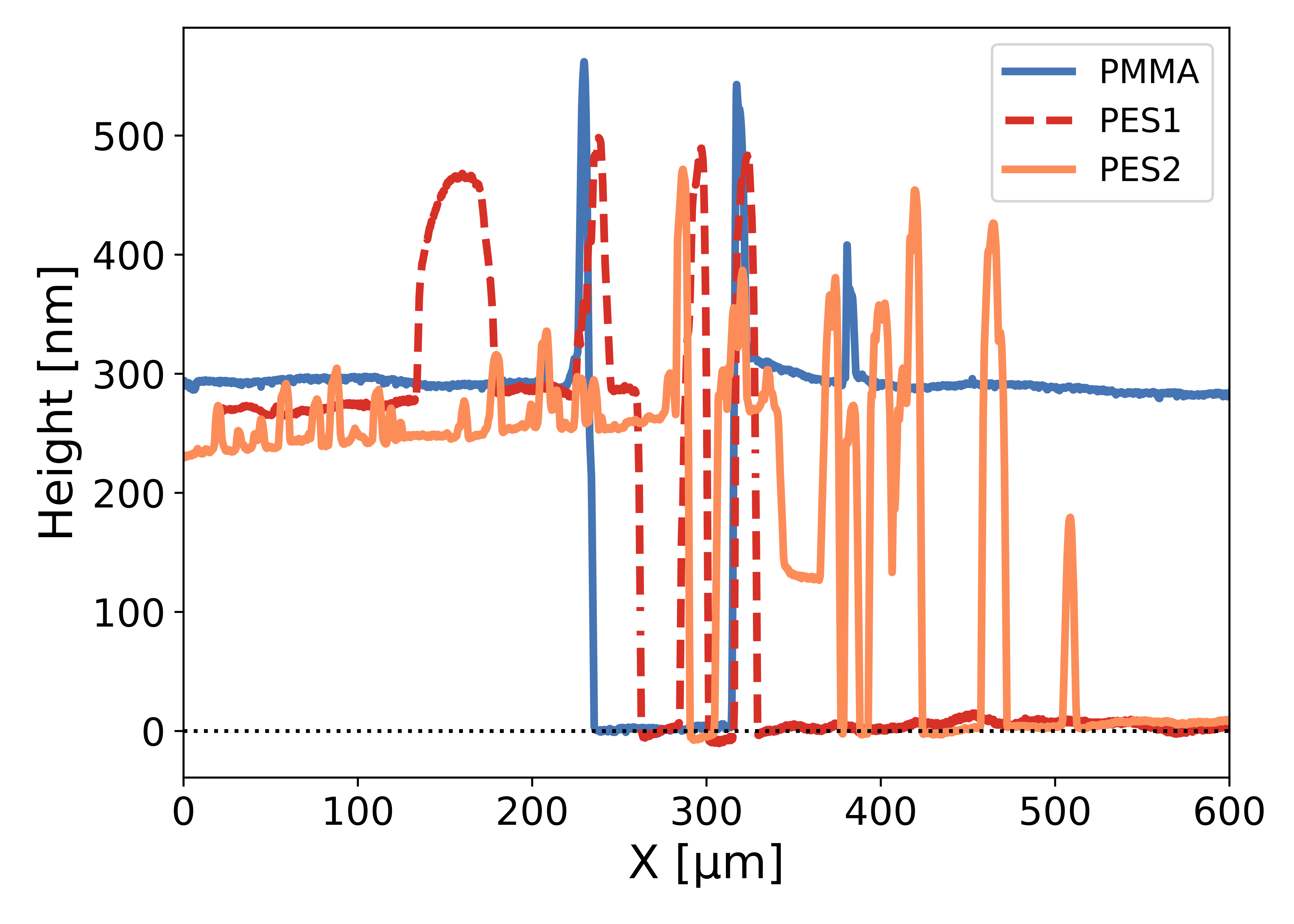}
    \caption{Thickness measurement of the thin films employed for the experiment: PMMA (blue solid line) and PES1 (red dashed line) are $\sim 290$ nm thick, while PES2 (orange solid line) is $\sim 270$ nm thick. Zero-height line is shown (black dotted line). The measurement was performed with a stylus profilometer.}
    \label{s_fig1}
\end{figure}

\begin{figure}
    \centering
    \includegraphics[width = .95\textwidth]{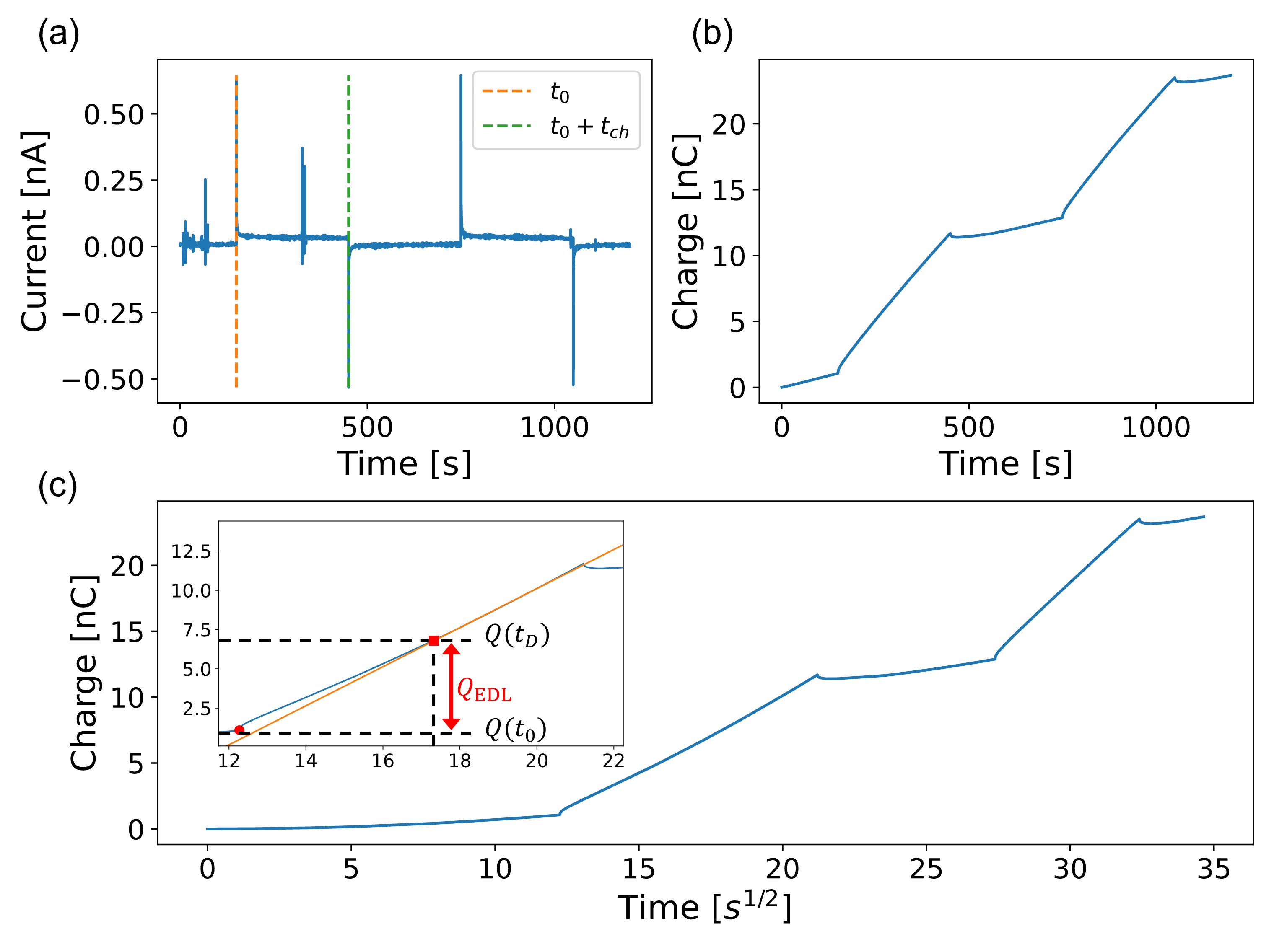}
    \caption{(a) Current over time measured for $V_{pp}$ = 300 mV in the two terminal setup described in the methods section of the main text. The voltage is applied at $t=t_0$ (orange dashed line) and is turned off at $t = t_0 + t_{ch}$ (green dashed line). (b) Charge moved during the measurement as a function of time. The curve is obtained by integrating the data shown in (a). (c) Charge moved during the measurement as a function of the squared time. In the inset, the method for extracting the charge accumulated during the first charging is shown. The linear part of the curve is fitted (orange solid line) and $t_D$ is determined as the time at which the curve enters the linear behaviour. The accumulated charge $Q_{EDL}$ is then extracted as the difference between the charge $Q(t_D)$ at time $t_D$ (red square in the picture) and the charge $Q(t_0)$ at time $t_0$ (red dot in the picture).}
    \label{s_fig2}
\end{figure}

\begin{figure}
    \centering
    \includegraphics[width = .9\textwidth]{Electrochemical windows.png}
    \caption{$I-V$ curves measured on a ES1 drop on metallic pads in air (red curve) and in vacuum (blue curve). $\Delta V_{EC}$ is the electrochemical windows of the liquids in the two conditions. At the onset of the electrochemical reactions, the ionic current changes, thus yielding a different conductivity in the system. The electrochemical windows measured in air is influenced by the interaction of the liquid with the water adsorbed from the atmosphere.}
    \label{s_fig3}
\end{figure}

\begin{figure}
    \centering
    \includegraphics[width = .9\textwidth]{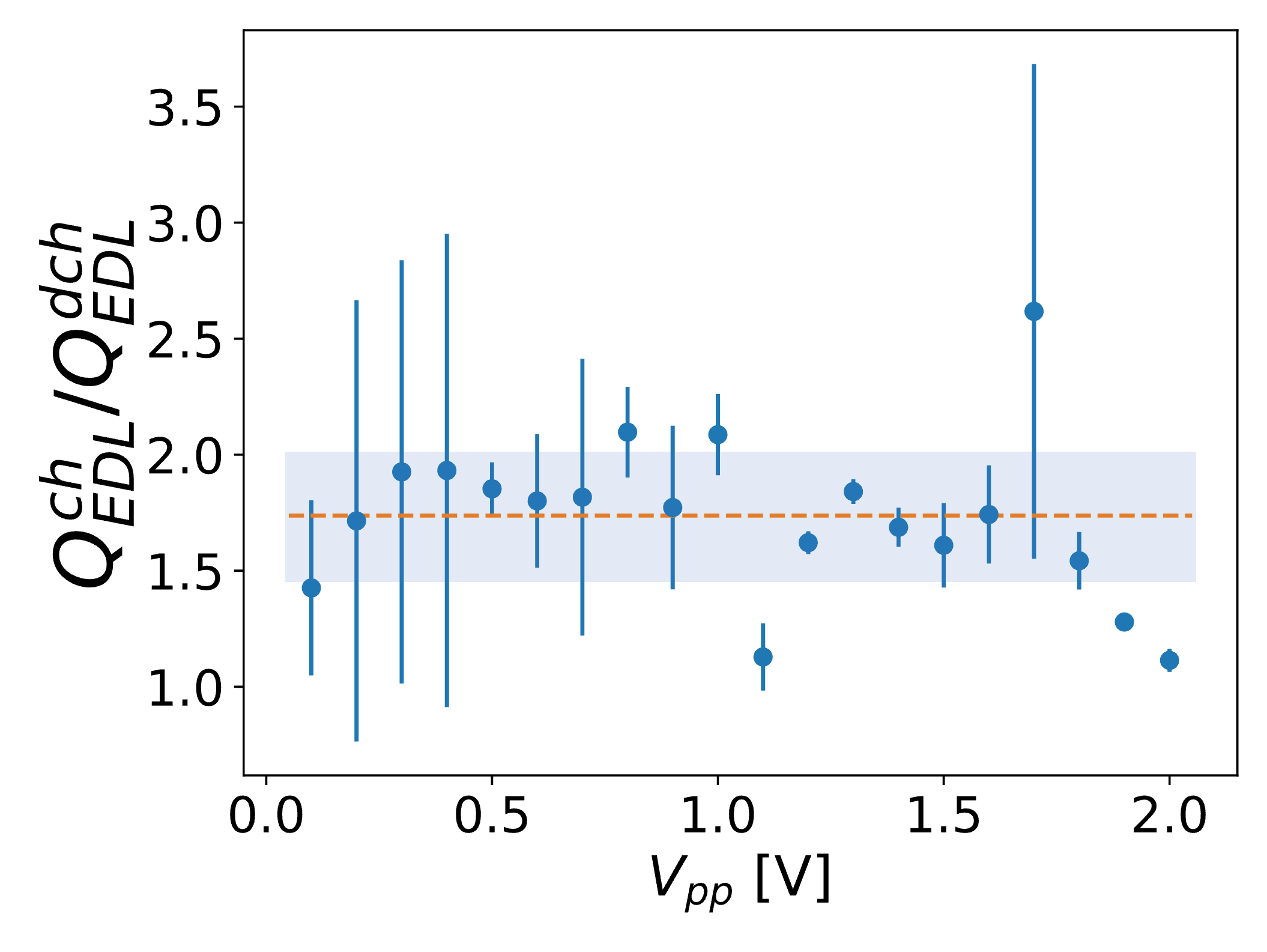}
    \caption{Charging/discharging ratio of the charge densities shown in Fig. 1b in the main text. The mean (orange dashed line) is 1.7 and the standard deviation (blue shaded area) is 0.3. The maximal point has a value of 2.6 for $V_{pp}$ = 1.7 V.}
    \label{s_fig4}
\end{figure}

\begin{figure}
    \centering
    \includegraphics[width = .9\textwidth]{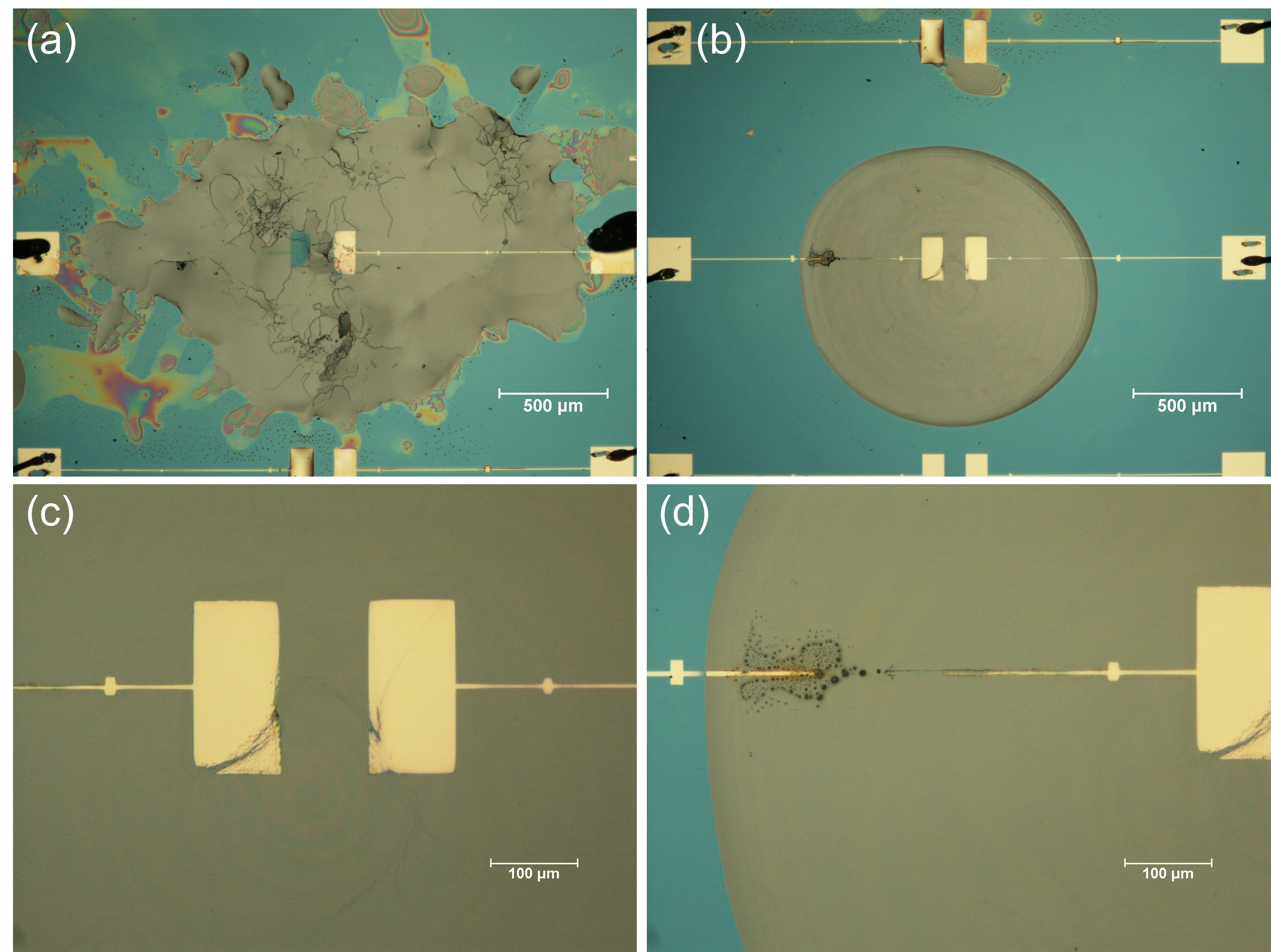}
    \caption{(a) Optical image of the device covered with ES1 after measurements in air. The acid attack is clearly visible, as one of the electrode is missing. The high end of the source meter unit was connected to the left electrode, while the right one was connected to the ground. (b) Optical image of the device covered with PES1 after measurements in air. (c) Close up image of the electrodes of the device shown in (b). Some modifications are visible due to the acid attack. (d) Close up of the high end electrode wire of the device shown in (b). Again, gold was removed from parts of the electrode from the bromide acid environment.}
    \label{s_fig5}
\end{figure}
\begin{figure}
    \centering
    \includegraphics[width = .9\textwidth]{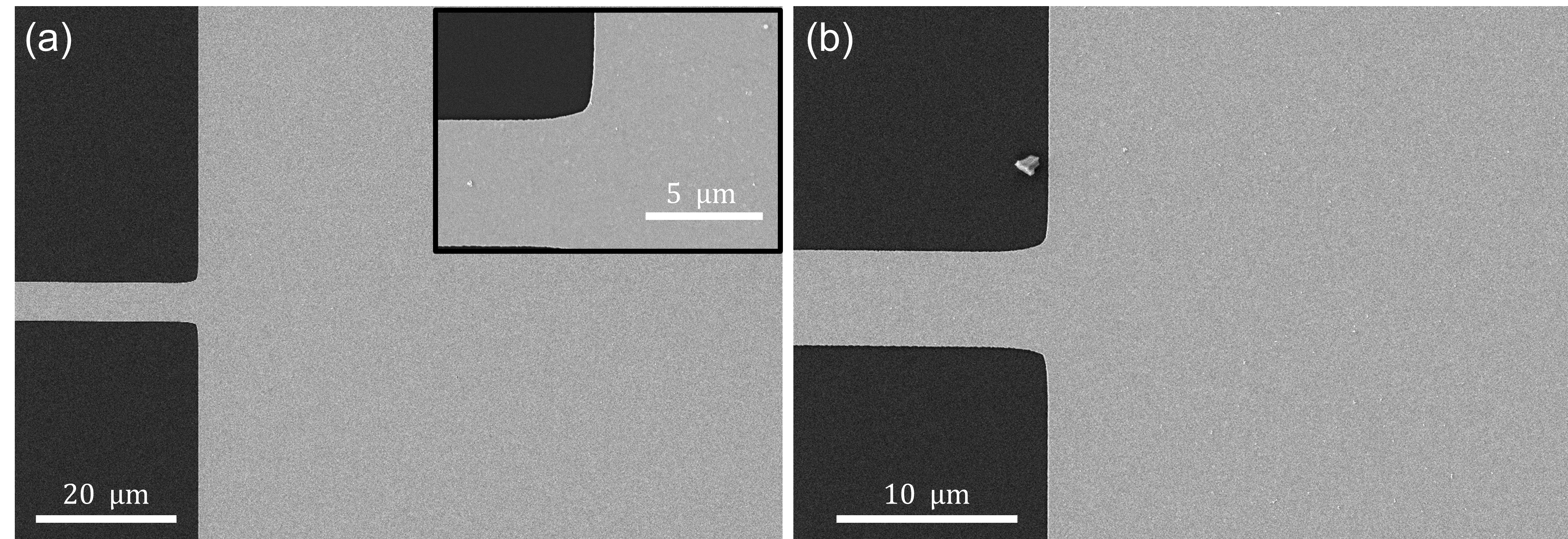}
    \caption{(a) SEM image of one electrode on which ES1 was deposited and left for 48 hours without the application of any voltage. In the inset, a close up of the region connecting the pad to the wire shows no sign of chemical attacks from bromide. (b) SEM image of the same electrode as in (a) prior to the application of ES1.}
    \label{s_fig6}
\end{figure}
--%
%